# A New Model for Black Hole Soft X-ray Transients in Quiescence

Ramesh Narayan, Jeffrey E. McClintock, & Insu Yi

Harvard-Smithsonian Center for Astrophysics
60 Garden Street, Cambridge, MA 02138;
jem@cfa.harvard.edu, rnarayan@cfa.harvard.edu, iyi@cfa.harvard.edu

## Abstract

We present models of the soft X-ray transients, A0620-00, V404 Cyg, and X-ray Nova Mus 1991, in quiescence. In each source, we postulate that there is an outer region, extending outward from about 3000 Schwarzschild radii, where the accretion flow is in the form of a standard thin disk. The outer disk produces most of the radiation we observe in the infrared, optical and UV bands. We propose that the disk undergoes an instability at its inner edge, perhaps by the mechanism described recently by Meyer & Meyer-Hofmeister for cataclysmic variables. The accreting gas is thereby converted into a hot corona which flows into the black hole as a nearly virial two-temperature flow. We describe the hot inner flow by means of a recently discovered stable solution of optically thin advection-dominated accretion. In this flow, most of the thermal energy released by viscous dissipation is advected into the black hole and only a small fraction, $\sim 10^{-4} - 10^{-3}$, of the energy is radiated. The radiation is in the form of Comptonized synchrotron and bremsstrahlung emission, and has a broad spectrum extending from optical to soft gamma-rays. The models we present are consistent with all the available data in the three sources. In particular, the X-ray emission from the hot inner flow fits the observed flux and spectral index of A0620-00. We derive a mass accretion rate of $\sim 10^{-11} M_\odot \text{yr}^{-1}$ in A0620-00 and Nova Mus, and $\sim$ few $\times 10^{-10} M_\odot \text{yr}^{-1}$ in V404 Cyg. The best fit to the data is obtained for a viscosity parameter $\alpha \sim 0.1 - 0.3$ in the hot flow. The models predict that all three sources must have substantial flux in hard X-rays and soft $\gamma$-rays. This prediction is testable in the case of V404 Cyg with current instruments. A necessary feature of our proposal is that most of the viscous energy released in the accretion is advected into the black hole without being radiated. The success of the models in fitting the observations therefore indicates that the accreting stars have event horizons and therefore confirms the black hole nature of these stars.

Subject headings: accretion — accretion disks — black hole physics — stars: binaries — X-rays: stars

# 1. Introduction

Soft X-ray transients (SXTs) are mass transfer binaries in which the accreting star is most often a black hole and the mass-transferring secondary is a low mass main sequence star (van Paradijs & McClintock 1995). SXTs are also known as X-ray novae, and both names refer to the fact that these sources occasionally have outbursts during which their optical and X-ray luminosities increase by several orders of magnitude, implying an enormous increase of mass accretion onto the central star. Two competing models have been proposed to explain the outbursts (see Lasota 1995 for a review). In the disk instability model (e.g. Huang & Wheeler 1989, Mineshige & Wheeler 1989), the mass transfer through the disk undergoes highly nonlinear variations because of a thermal instability, while in the competing mass transfer instability model (e.g. Hameury, King & Lasota 1986, 1990, Goutitakis & Hameury 1993), the mass accretion rate from the secondary itself varies as a result of X-ray irradiation. The disk instability model has been successful in explaining several features of SXT outbursts and is currently favored (e.g. Lasota 1995). However, neither model has been able to explain the quiescent state of SXTs satisfactorily.

The problem of modeling SXTs in quiescence has become more acute with the detection of X-ray emission in the prototypical SXT, A0620-00 (McClintock et al. 1995). The optical luminosity of A0620-00 suggests a mass accretion rate $\dot{M} \sim 10^{-10} M_\odot \mathrm{yr}^{-1}$ in the outer parts of the accretion disk, while the observed X-ray luminosity implies $\dot{M} \sim 2 \times 10^{-15} M_\odot \mathrm{yr}^{-1}$ in the inner regions of the disk, if one assumes that all the gravitational energy released by accretion is radiated in X-rays (McClintock, Horne, & Remillard 1995). In the disk instability model, the maximum $\dot{M}$ in the inner disk during quiescence depends on the viscosity parameter $\alpha$. If $\alpha$ has a reasonable value on the order of $\sim 10^{-2} - 10^{-1}$, then there should be virtually no mass transfer in the inner disk and the X-ray flux should be many orders of magnitude less than what is observed (e.g. Lasota 1995). The X-ray luminosity can be explained only by adjusting $\alpha$ to be unusually low, $\alpha < 10^{-5}$. However, even if we set $\alpha$ to such a low value, or if we give up the disk instability model altogether and just allow $\dot{M}$ in the inner disk to be an adjustable parameter, we are still faced with a problem. At an accretion rate of $\dot{M} \sim 2 \times 10^{-15} M_\odot \mathrm{yr}^{-1}$, a standard thin accretion disk will be too cool to radiate at a temperature of 0.2 keV as observed in A0620-00. Thus, it is impossible to fit both the X-ray luminosity and the X-ray spectrum of A0620-00 with any standard thin disk model.

Assuming that the X-rays are produced close to the accreting black hole and that the optical emission comes from farther out, the data suggest that the gas near the center must be very hot. Further, the X-ray luminosity is less than the optical luminosity, whereas in a normal accretion flow with a constant mass accretion rate we would expect the X-ray luminosity to be much greater since most of the gravitational potential energy of the accreting gas is released near the black hole. Therefore, we deduce that either the mass accretion rate at the X-ray-emitting radii is much smaller than at the optical radii or that the radiative efficiency of the hot material is much less than that of the optical-emitting gas.

In this paper, we propose a new model for the quiescent state of SXTs which provides a natural explanation of the optical and X-ray data. Our model is based on the idea that the accretion flow in quiescent SXTs has two distinct zones: (i) a standard thin disk at large radii which produces the optical and UV radiation, and (ii) a hot quasi-spherical flow in the inner regions which produces the X-rays. The hot inner flow corresponds to an optically thin solution of the accretion equations discovered recently by Narayan & Yi (1995b) and Abramowicz et al. (1995) (see also the discussion of Rees et al. 1982 on the related "ion torus" model). This solution is viscously and thermally stable, in contrast to other hot accretion flow solutions discussed previously in the literature (Shapiro, Lightman & Eardley 1976, Kusunose & Takahara 1989, Wandel & Liang 1991). The flow is moreover *advection-dominated*, which means that most of the viscously dissipated energy is carried with the accreting gas as internal energy rather than being radiated.

A feature of our model is that it has essentially the same mass accretion rate in the outer and inner flows. The outer thin disk radiates efficiently, and so its luminosity is given by $L_{out} \sim GM\dot{M}/R_{in}$, where $M$ is the mass of the accreting black hole and $R_{in}$ is the inner radius of the outer disk. The hot inner flow on the other hand is advection-dominated and so radiates much less than the nominal luminosity $\sim 0.1 GM\dot{M}/R_S$ which we would normally expect from an accretion flow, where $R_S$ is the Schwarzschild radius of the black hole. In fact, the radiative efficiency of our inner flow is only $\sim 10^{-4} - 10^{-3}$, i.e. $L \sim (10^{-4} - 10^{-3}) GM\dot{M}/R_S$, and this explains the low X-ray luminosity observed in these sources.

The paper is organized as follows. In §2, we describe the model, explaining in some detail the advection-dominated inner flow and the various components which contribute to the spectrum. In §3, we fit the data on A0620-00 in quiescence and show that our model explains the observations quite well. We also come up with constraints on some of the parameters such as the viscosity parameter $\alpha$ and the transition radius $R_{in}$. In §4, we apply our model to two other SXTs, V404 Cyg and X-ray Nova Muscae 1991 (hereafter Nova Mus). There are fewer data constraints on these systems and so we are able to fit the data both with our model and with a standard steady state thin disk. We point out key future observations which will be able to distinguish between the two classes of models. We conclude with a discussion in §5.

## 2. Description of the Model

### 2.1. The Hot Advection-Dominated Inner Flow

We model the hot accretion flow in the inner regions of a quiescent SXT in terms of the new optically thin advection-dominated solutions described recently in several papers (Narayan & Yi 1994, 1995ab, Abramowicz et al. 1995, Chen 1995, Chen et al. 1995, see also Rees et al. 1982). Because the accreting material in this solution is optically thin, its cooling is very inefficient. Consequently, most of the accretion energy is stored in the gas as internal energy and the ions in the accreting gas heat up almost to virial temperatures. The flow is thus quasi-spherical in morphology (Narayan & Yi 1995a), with sub-Keplerian rotation and a relatively large radial velocity (see eqs 2.5, 2.6 below).

The models of SXTs presented in this paper are based on a particular version of advection-dominated flows which we have developed in previous papers (Narayan & Yi 1994, 1995b). We assume a steady axisymmetric flow and average the flow equations vertically so that we reduce the problem to a two-dimensional flow in the equatorial $R\phi$ plane. We have shown (Narayan & Yi 1995a) that the vertical averaging approximation is quite good even for nearly spherical flows. We write the pressure as $p = \rho c_s^2$, where $\rho(R)$ is the height-averaged density at radius $R$ and $c_s(R)$ is the isothermal sound speed. As we explain below, the pressure is partly due to the gas and partly due to magnetic fields. We denote the Keplerian angular velocity by $\Omega_K(R) = (GM/R^3)^{1/2}$, the vertical scale height of the gas as $H = c_s/\Omega_K$, and write the kinematic coefficient of shear viscosity as $\nu = \alpha c_s H = \alpha c_s^2/\Omega_K$, where $\alpha$ is a constant (Shakura & Sunyaev 1973).

With the above definitions, the density of the gas $\rho$, the radial velocity $v$, angular velocity $\Omega$, and isothermal sound speed $c_s$, satisfy the following four differential equations (cf. Narayan & Yi 1994):

$$\frac{d}{dR}(\rho R H v) = 0, \tag{2.1}$$

$$v\frac{dv}{dR} - \Omega^2 R = -\Omega_K^2 R - \frac{1}{\rho}\frac{d}{dR}(\rho c_s^2), \tag{2.2}$$

$$\frac{v}{R^2}\frac{d(\Omega R^2)}{dR} = \frac{1}{\rho R^3 H}\frac{d}{dR}\left(\frac{\alpha \rho c_s^2 R^3 H}{\Omega_K}\frac{d\Omega}{dR}\right), \tag{2.3}$$

$$\frac{3+3\epsilon}{2} 2\rho H v \frac{dc_s^2}{dR} - 2c_s^2 v \frac{d(\rho H)}{dR} = f \frac{2\alpha \rho c_s^2 R^2 H}{\Omega_K}\left(\frac{d\Omega}{dR}\right)^2. \tag{2.4}$$

In the final equation, the left hand side gives the rate of change of entropy of the accreting gas, while the right hand side is $f$ times the rate of energy dissipation through viscosity. The parameter $f$, which is in general a function of $R$, describes the fraction of the energy which is stored in the gas. In the solutions presented here, the radiative efficiency $1-f$ is very much less than unity, so $f \to 1$ and the flows are highly advection-dominated. For convenience, we have defined $\epsilon = (5/3 - \gamma)/(\gamma - 1)$, where $\gamma$ is the ratio of specific heats.

Since $f$ is practically equal to unity, it is effectively independent of $R$. When $f$ is a constant, equations (2.1)–(2.4) have a self-similar solution (Narayan & Yi 1994, Spruit et al. 1987) of the form

$$v(R) = -c_1 \alpha v_{ff}, \tag{2.5}$$

$$\Omega(R) = c_2 v_{ff}/R, \tag{2.6}$$

$$c_s^2(R) = c_3 v_{ff}^2, \tag{2.7}$$

where $v_{ff}(R) = R\Omega_K(R)$ is the free-fall velocity at radius $R$, and

$$c_1 = \frac{5 + 2\epsilon'}{3\alpha^2} g(\alpha, \epsilon'), \tag{2.8}$$

$$c_2 = \left[\frac{2\epsilon'(5 + 2\epsilon')}{9\alpha^2} g(\alpha, \epsilon')\right]^{1/2}, \tag{2.9}$$

$$c_3 = \frac{2(5 + 2\epsilon')}{9\alpha^2} g(\alpha, \epsilon'), \tag{2.10}$$

$$g(\alpha, \epsilon') \equiv \left[1 + \frac{18\alpha^2}{(5 + 2\epsilon')^2}\right]^{1/2} - 1, \tag{2.11}$$

with $\epsilon' = \epsilon/f$. We use this self-similar solution throughout the present work. There are two main approximations in this approach. First, we assume a purely Newtonian form for the gravitational potential instead of using the appropriate relativistic metric close to the black hole. We do not expect this to be a serious source of error. Secondly, we assume that the self-similar solution is valid throughout the hot flow. Close to the boundaries we expect the flow in fact to deviate from self-similarity. At the outer edge of the hot flow, the rotation has to match the Keplerian rotation of the thin disk. Narayan & Yi (1994) show that the exact solution corresponding to this boundary condition relaxes to the self-similar form within a short distance from the boundary. Similarly, we expect a sonic point to be present at small radii so that the flow would accrete supersonically onto the black hole. Therefore, in the very innermost region the flow will not be self-similar, but we expect the self-similar solution to be be valid for radii beyond a factor of a few of the sonic radius. Thus, the solution we employ is likely to be a good description of the bulk of the accreting gas.

We assume that the accreting gas in the hot flow contains magnetic fields in quasi-equipartition with the gas, such that a constant fraction $(1 - \beta)$ of the pressure is due to the fields. The rest of the pressure is contributed by the gas. Because the gas is highly optically thin, radiation pressure is negligible and so we do not include it. The ratio of specific heats is then given by $\gamma = (32 - 24\beta - 3\beta^2)/(24 - 21\beta)$ (Narayan & Yi 1995b). Following Shapiro et al. (1976) and Rees et al. (1982), we assume that the accreting gas is a two-temperature plasma, where the ion temperature $T_i$ and electron temperature $T_e$ are allowed to be different from each other. We thus write the gas pressure as

$$p_g = \beta\rho c_s^2 = \frac{\rho k T_i}{\mu_i m_u} + \frac{\rho k T_e}{\mu_e m_u}, \tag{2.12}$$

and take the effective molecular weights of the ions and electrons to be $\mu_i = 1.23$, $\mu_e = 1.14$ (Narayan & Yi 1995b).

We scale the black hole mass in solar mass units,

$$M = mM_\odot, \tag{2.13}$$

and the mass accretion rate in Eddington units,

$$\dot{M} = \dot{m}\dot{M}_E, \qquad \dot{M}_E = \frac{4\pi GM}{0.1\kappa_{es}c} = 1.39 \times 10^{18}m \qquad (2.14)$$

where $\kappa_{es} = 0.4\text{cm}^2\text{g}^{-1}$ is the opacity due to electron scattering. We scale radii in Schwarzschild gravitational units,

$$R = rR_S, \qquad R_S = \frac{2GM}{c^2} = 2.95 \times 10^5 m \text{ cm.} \qquad (2.15)$$

In equation (2.14), we have followed the standard practice (e.g. Frank et al. 1992) of assuming a nominal efficiency of 10% to calculate the Eddington accretion rate. However, this is purely a matter of definition. In advection-dominated flows around black holes such as the ones we consider in this paper, the actual radiation efficiency is very low since the black hole swallows all the advected energy. It is this effect which enables our models to explain the very low X-ray luminosities of quiescent SXTs.

In the present work we set $\beta = 0.95$, corresponding to a magnetic pressure that is 5% of the total pressure. (Equipartition, $\beta = 0.5$, assumes that one half of the total pressure is contributed by the magnetic fields.) As we have shown in Narayan & Yi (1995b), other values of $\beta$ lead to only minor variations in the results. (For instance, the quantitative results in the quoted paper changed by only about 10% when $\beta$ was varied between 0.5 and 0.95.) Let us for the moment also set $f \approx 1$ (because of advection-domination) and assume $\alpha^2 \ll 1$. We then obtain $\gamma = 1.603$, $\epsilon' \approx \epsilon = 0.106$, and $c_1 = 0.576$, $c_2 = 0.206$, $c_3 = 0.384$. We can then write the following approximate scaling laws for the various physical quantities in the flow (see Narayan & Yi 1995b for details)

$$v = -1.22 \times 10^{10} \alpha r^{-1/2} \text{ cm s}^{-1}$$
$$\Omega = 1.48 \times 10^4 m^{-1} r^{-3/2} \text{ s}^{-1},$$
$$c_s^2 = 1.73 \times 10^{20} r^{-1} \text{ cm}^2\text{s}^{-2},$$
$$\rho = 1.06 \times 10^{-4} \alpha^{-1} m^{-1} \dot{m} r^{-3/2} \text{ g cm}^{-3},$$
$$p = 1.84 \times 10^{16} \alpha^{-1} m^{-1} \dot{m} r^{-5/2} \text{ g cm}^{-1}\text{s}^{-2},$$
$$B = 1.52 \times 10^8 \alpha^{-1/2} m^{-1/2} \dot{m}^{1/2} r^{-5/4} \text{ G},$$
$$n_e = \rho/\mu_e m_u = 5.60 \times 10^{19} \alpha^{-1} m^{-1} \dot{m} r^{-3/2} \text{ cm}^{-3},$$
$$\tau_{es} = 2n_e \sigma_T H = 21.7 \alpha^{-1} \dot{m} r^{-3/2}$$
$$q^+ = 1.21 \times 10^{21} m^{-2} \dot{m} r^{-4} \text{ erg cm}^{-3}\text{s}^{-1}. \qquad (2.16)$$

Here, $B$ is the magnetic field strength, $n_e$ is the electron number density, $\tau_{es}$ is the scattering optical depth, $\sigma_T = 6.62 \times 10^{-25}$ cm$^2$ is the Thomson cross-section, and $q^+$ is the volume rate of dissipation of energy by viscosity. Finally, by using $p_g = \beta p$ in equation (2.12), we obtain a relation connecting $T_i$ and $T_e$ at each radius,

$$T_i + 1.08 T_e = 2.43 \times 10^{12} r^{-1} \text{ K}. \qquad (2.17)$$

Note that we have set $f = 1$ in the above relations just in order to write down explicit estimates of the physical variables. For the detailed calculations presented later, however, we solve for $f$ as a function of $r$ self-consistently in the manner described in Narayan & Yi (1995b). Therefore, the solutions differ from the above scalings by small factors (small because in practice $f$ is very nearly equal to 1 in all cases).

We determine the ion and electron temperatures in the accreting plasma by taking into account the detailed balance of heating, cooling, and advection. Due to the large mass difference between ions and electrons, we expect the viscous heating to act mainly on the ions (Shapiro et al. 1976, Rees et al. 1982). The ions then transfer some of their energy to the electrons through Coulomb coupling at a volume rate $q^{ie}$ which can be calculated as a function of $n_e$, $T_i$ and $T_e$. In principle, the energy transfer could also occur by other collective processes (e.g. Begelman & Chiueh 1988) but we ignore this possibility (see Appendix A in Narayan & Yi 1995b), as in previous work on two-temperature accretion flows (Shapiro et al. 1976).

The cooling of the plasma is primarily via electrons and occurs through a variety of channels such as bremsstrahlung (at a volume rate $q_{br}^-$), Comptonization of the bremsstrahlung photons ($q_{br,C}^-$), synchrotron emission ($q_{synch}^-$), and its Comptonization ($q_{synch,C}^-$). The total cooling rate is the sum of the individual rates: $q_{tot}^- = q_{br}^- + q_{br,C}^- + q_{synch}^- + q_{synch,C}^-$. Estimates of the individual $q$'s involve the microphysics of the radiating plasma; details are given in Narayan & Yi (1995b).

Using the above volume rates, we close the equations with two additional relations which ensure the thermal energy balance of the ions and the electrons. For the ions, we require the rate of input of energy through viscous dissipation to be equal to the sum of the rate of advection of energy by the ions and the rate of transfer of energy from the ions to the electrons. This gives

$$q^+ = q^{adv} + q^{ie} = fq^+ + q^{ie}. \qquad (2.18)$$

We solve this equation at each radius to obtain a self-consistent value of $f(r)$. In practice, $q^{ie} \ll q^+$ in all cases and $f$ is very close to unity. The energy balance of the electrons provides a second condition. For thermal balance, the Coulomb heating rate of the electrons and the radiative cooling rate should be equal, i.e.

$$q^{ie} = q^-. \qquad (2.19)$$

This equation provides the final relation which enables us to solve for $T_i$ and $T_e$ at each radius of the flow. In combination with equations (2.16)–(2.18), we thus obtain a complete solution for all the physical variables in the inner advection-dominated flow.

Since we have estimates of the masses $M$ of the accreting black holes in the various SXTs we model, and since we set $\beta = 0.95$, our model of the inner flow has only two adjustable parameters, the mass accretion rate $\dot{M}$ and the viscosity parameter $\alpha$. Actually, there is a degeneracy even between $\dot{M}$ and $\alpha$, and the emission from a given model depends to a very good approximation only on the particular combination $\dot{M}/\alpha$. Thus, given measurements of the X-ray flux from a quiescent SXT, our model enables us to determine $\dot{M}/\alpha$ immediately.

To conclude this subsection, we would like to emphasize that although the advection-dominated, optically-thin solution employed in this work is relatively new in the literature, it describes a perfectly well-behaved and physically consistent accretion flow. The flow differs considerably from the standard thin accretion disk in that it is not geometrically thin, nor is it Keplerian or cooling-dominated. Nevertheless, the solution describes a viable accretion flow, which is both thermally and viscously stable. The flow is, however, convectively unstable (Narayan & Yi 1994, 1995a), and it is possible that convection may have the effect of increasing the viscosity of the accreting gas.

Another point to emphasize is that the advection-dominated flows we consider here are different from the hot two-temperature flows described by Shapiro et al. (1976) and later authors (e.g. Kusunose & Takahara 1989, Wandel & Liang 1991). The Shapiro et al. solutions are cooling-dominated (i.e. $f \ll 1$) and are thermally unstable (Piran 1978). Our two-temperature flows are hotter and less dense at a given $\dot{M}$ than the Shapiro et al. solutions, but avoid the thermal instability through advection (Abramowicz et al. 1995, Narayan & Yi 1995b).

## 2.2 The Outer Thin Accretion Disk

Our model of the flow at large radii corresponds to a standard thin accretion disk (Shakura & Sunyaev 1973, Novikov & Thorne 1973, Lynden-Bell & Pringle 1974, Frank, King & Raine 1992), except that we truncate the disk at an inner radius $R_{in}$ which is well away from the black hole. We assume that the disk is in steady state and that it is optically thick. The emission is then blackbody at each radius. The luminosity and spectrum are determined by the mass accretion rate $\dot{M}(=\dot{m}M_E)$, the inner radius $R_{in}(=r_{in}R_S)$ of the disk, and the outer radius $R_{out}(=r_{out}R_S)$. (Throughout this paper, we use the symbol $R$ to represent the radius in physical units and $r$ for the radius in Schwarzschild units, cf. eq. 2.15.) The viscosity parameter $\alpha_t$ within the thin disk does not influence the emission so long as the disk is optically thick.

In our models, we assume that $\dot{M}$ in the outer disk is the same as that in the inner hot flow. Since most of the disk emission occurs near its inner edge, the scaled outer radius $r_{out}$ affects the results only weakly. We determine $r_{out}$ in the individual objects following the procedure described by Smak (1981), where $r_{out}$ corresponds to the radius at which the projected Keplerian velocity is equal to half the peak separation in the H$\alpha$ profile (e.g. Horne & Marsh 1986). In principle, we could apply a similar procedure for the inner radius $r_{in}$ as well and compute its value from the largest velocity observed in the H$\alpha$ profile. However, since the velocity at which the H$\alpha$ line goes to zero intensity is often difficult to determine, we have used $r_{in}$ as a free parameter of the model.

In our models, the optical flux is generated primarily by the outer thin disk. The net luminosity of this region of the flow is given by $L_{out} \sim GM\dot{M}/R_{in}$. Thus, the observed optical luminosities constrain $\dot{M}/R_{in}$ in each source. In addition, the characteristic temperature of the optical emission depends on $M\dot{M}/R_{in}^3$ (Frank et al. 1992). Thus, if the optical data provide a reliable estimate of the maximum temperature in the disk, then we can determine both $\dot{M}$ and $R_{in}$ to within factors of a few.

The critical radius $R_{in}$ plays an important role in our model since it separates the efficiently radiating cool outer disk from the inefficiently radiating hot inner flow. We treat $R_{in}$ essentially as a fitting parameter which we adjust so as to achieve good agreement between the model spectrum and the data. It is clearly important to study the physics of the transition between the two kinds of flows and to calculate $R_{in}$ from first principles. In this context we note that Meyer & Meyer-Hofmeister (1994) have recently proposed a "siphon flow" mechanism to explain some properties of accretion disks in low-$\dot{M}$ cataclysmic variables. In their model, the accretion occurs initially via a thin disk at large radii from the white dwarf, but the material in the disk continuously evaporates from the surface because of thermal conduction of energy from a corona. As a result, the disk becomes truncated at a particular radius and the accretion at smaller radii occurs via a hot flow. The Meyer & Meyer-Hofmeister mechanism appears to be very promising for SXTs as well, and if it were to operate in these sources would give precisely the kind of two-zone model we propose. It remains to be seen if the model can quantitatively reproduce the critical radii $R_{in}$ which we derive in this paper by fitting the data.

For simplicity, we have assumed that the mass accretion rate $\dot{M}$ is independent of radius in the outer thin disk of our model. In the disk instability model of SXT outbursts (Huang & Wheeler 1989, Mineshige & Wheeler 1989), $\dot{M}$ is expected to vary with $R$ during the quiescent phase. Since we do not have a self-consistent model of the outburst we feel it is better to assume $\dot{M}$ to be constant than to include an arbitrary dependence on $R$.

### 2.3. Calculation of the Spectrum

We take the effective blackbody temperature of the outer thin disk to be given by (cf. Frank et al. 1992)

$$T_{eff}(R) = \left[\frac{3GM\dot{M}}{8\pi\sigma R^3}\right]^{1/4} = 6.3 \times 10^7 m^{-1/4}\dot{m}^{1/4}r^{-3/4}, \qquad (2.20)$$

where $\sigma$ is the Stefan-Boltzmann constant. The flux we observe at distance $D$ from the source is then obtained by integrating the blackbody emission over radius:

$$F_\nu d\nu = \frac{4\pi h\nu^3 \cos i \, d\nu}{c^2 D^2} \int_{R_{in}}^{R_{out}} \frac{R dR}{[\exp(h\nu/kT_{eff}(R)) - 1]}. \qquad (2.21)$$

Here $i$ is the inclination of the disk, which we take to be the same as the inclination of the orbital plane of the binary system. We obtain $M$, $D$ and $i$ from the observations.

In the inner region, the emission is from an optically thin hot magnetized plasma where the density, temperature, magnetic field strength, etc. all vary with radius. The emission from this medium involves several processes and needs to be calculated in detail. Fortunately, we have two simplifying features. First, the flow is nearly spherical and so we can integrate over spherical shells. Second, the flow is optically very thin and so we negelct all of the radiative transfer effects except for synchrotron self-absorption (Narayan and Yi 1995b).

We calculate the emission shell by shell, starting from the innermost radius close to the black hole horizon, and integrating outwards. Going from one shell to the next, we use the local electron density and temperature to calculate the effect of Comptonization on the local radiation. We also apply the appropriate gravitational redshift between the two radii. Then, at the new shell, we calculate the emission due to bremsstrahlung and synchrotron processes, allowing for synchrotron self-absorption which is very important (see Narayan & Yi 1995b for details). We add the bremsstrahlung and synchrotron emission to the radiation already present, and then propagate the combined radiation to the next shell, with Comptonization and redshift as before. We ignore Compton-cooling by soft photons emitted by the cool outer disk. This is justified in these models because the energy density due to disk photons in the relevant region of the flow is always very much less than the energy density due to the local synchrotron radiation.

For the Comptonization, we assume that the electrons have large Lorentz $\gamma$ (e.g. Blumenthal & Gould 1970) and make use of the appropriate simplification. We include the Klein-Nishina reduction of the cross-section at large photon energies and we also allow for saturation of the Comptonization when $h\nu \to kT_e$. We calculate the bremsstrahlung emission using standard results appropriate to a relativistic plasma (Svensson 1982, Stepney & Guilbert 1983), using an analytical expression for the variation of the Gaunt factor with $\nu$ (Novikov & Thorne 1973). Finally, we calculate the self-absorbed synchrotron emission using fitting formulae given in Narayan & Yi (1995b).

The typical spectrum we obtain with our model of the hot inner flow consists of three peaks in $\nu F_\nu$ (see the dashed line in Fig. 1). There is a peak at optical frequencies due to synchrotron emission, which is usually weaker than the optical blackbody emission from the outer thin disk. The Comptonization of the synchrotron photons leads to a second peak in soft X-rays which in the case of A0620-00 fits the observed X-ray flux very well. Finally, there is a third extended peak in hard X-rays which is a combination of multiply Compton scattered synchrotron photons and bremsstrahlung photons. In principle, there are also Comptonized bremsstrahlung photons in this peak, but their contribution is small.

## 3. Modeling the Soft X-Ray Transient A0620-00

### 3.1 Data for A0620-00

We first select a set of system parameters for A0620-00 that we use as inputs to our models for the source. These include the black hole mass, the mass accretion rate, the distance, and the velocities at the inner and outer edges of the accretion disk. Secondly, we summarize the multiwavelength data (X-ray, EUV, UV, optical, IR and radio) that we use to constrain our models.

A0620-00 (=Nova Mon 1917, 1975) is the most thoroughly studied X-ray nova and it has the shortest confirmed orbital period, P = 7.8 hr. The starting place for developing a model is the value of the mass function (Orosz et al. 1994):

$$f(M) \equiv \frac{PK_s^3}{2\pi G} = \frac{(M \sin i)^3}{(M + M_s)^2} = 2.91 \pm 0.08 M_\odot, \quad (3.1)$$

where $K_s$ is the semiamplitude of the velocity curve of the secondary star, $i$ is the orbital inclination angle, $M$ is the mass of the black hole primary, and $M_s$ is the mass of the K-dwarf secondary.

The mass of the black hole depends weakly on the mass of the secondary star, which we fix at $M_s = 0.5~M_\odot$. The crucial parameter is the inclination angle, $i$, and we consider two very different values which are suggested by observation: $i = 70°$ (Haswell et al. 1993) and $i = 40°$ (Shahbaz, Naylor, and Charles 1994). The corresponding black hole masses are $M = 4.4~M_\odot$ and $M = 12~M_\odot$, respectively.

Dynamical and geometrical information about the accretion disk can be obtained from studies of the broad, double-peaked Balmer lines (Smak 1981; Horne & Marsh 1986). Of interest in the present study are $v_{out}$ and $v_{in}$, the projected velocities at the outer and inner edges of the accretion disk. Estimates of these velocities have been inferred from orbit-averaged profiles of the H$\alpha$ emission line by several groups. For the velocity at the outer edge we adopt the value derived by fitting Smak models to the line profiles: $v_{out} = 550 \pm 10$ km s$^{-1}$ (Johnston, Kulkarni, and Oke 1989; Orosz et al. 1994). The scaled outer radius of the disk is then given by

$$r_{out} = \frac{1}{2} \left( \frac{c \sin i}{v_{out}} \right)^2 = 1.3 \times 10^5. \quad (3.2)$$

The velocity at the inner edge is the half width at zero intensity of the H$\alpha$ profile corrected for instrumental resolution. The profiles plotted in Marsh et al. (1994) and Orosz et al. (1994) suggest the following value: $v_{in} = 2100$ km s$^{-1}$. This is actually a lower limit on $v_{in}$; higher velocity gas is present, although difficult to detect. For example, values of $v_{in}$ as high as about 2800 km s$^{-1}$ have been inferred (see Fig. 10 in Marsh et al. 1994). Thus, by (3.2), $r_{in} < 9000$.

An average value for the mass accumulation rate in the accretion disk can be obtained from the energy released in the 1975 outburst, $\sim 3\times10^{44}$ erg, and the 58 yr interval between outbursts. If one assumes a Schwarzschild black hole and a standard accretion efficiency of 0.1 during outburst, one obtains a quiescent mass storage rate of $\dot{M} \sim 3\times 10^{-11} M_\odot \text{yr}^{-1}$ (McClintock et al. 1983). This estimate is based on the assumption that no outbursts were missed between 1917 and 1975.

Oke (1977) estimated the distance to A0620-00 by assuming that the secondary is a main-sequence K5-K7 dwarf: $D = 870 \pm 200$ pc. A somewhat earlier spectral type, K4-5 (Haswell et al. 1993; McClintock et al. 1995), and less reddening, E(B-V) = 0.35 mag (Wu et al. 1983), are now favored, which implies $M_v$ = 7.0 - 7.4 (Lang 1991) and $D = 1200 - 1500$ pc (where we have adopted $A_v$/E(B-V) = 3.1). Of course, the secondary may not have the radius of a main sequence star. Shahbaz, Naylor and Charles (1994) used photometric data, a model for the system, and a relationship between the angular diameter and effective temperature of the secondary to obtain $D = 1050 \pm 400$ pc. In a similar manner, Marsh et al. (1994) derived the following distance estimates: D = 500 ($i = 70°$) and $D = 700$ pc ($i = 40°$). In this paper we adopt the value $D = 1.0$ kpc.

The multiwavelength data, $F_\nu$ vs. $\nu$, are summarized in Table 1. The X-ray flux (entry 1) is an average over the 0.4-2.4 keV band and was derived from ROSAT PSPC observations made in 1992 March, which are described in McClintock et al. (1995). In the X-ray data analysis, we followed precisely the steps described in McClintock et al. (1995) with the following three exceptions: (1) a power law model of the spectrum was used, which is more relevant to the present work; (2) a wider range of pulse heights was included (channels 10-34; nominally 0.4-2.5 keV); and (3) only the spectral index was varied in determining the 1 $\sigma$ errors.

As described in McClintock et al. (1995), only 39 $\pm$ 8 net source counts were detected in the $3 \times 10^4$ s observation. Consequently, we fixed the interstellar column density at $N_H = 1.6 \times 10^{21}$ cm$^{-2}$ and used maximum-likelihood fitting, which requires fitting the source and background regions simultaneously. The background spectrum was fitted with a simple power law model, which gives a better fit than blackbody or bremsstrahlung models. Only the spectral index of the source was varied in determining the uncertainties in that parameter; all of the other parameters (the source intensity, and the parameters of the background spectrum) were fixed at their best-fit values. Then, 300 sets of simulated pulse height data were generated and fitted to give the following value and (1 $\sigma$) uncertainties for the photon spectral index: $\alpha_N$ = 3.5 (+0.8,-0.7). The corresponding X-ray flux is log $F_\nu$ = -31.16 (-0.01,+0.04), and the luminosity is $L_X \approx 4 \times 10^{30}$ erg s$^{-1}$. It is probable that most of the X-ray flux comes from the vicinity of the black hole, and not from the secondary star, as discussed in McClintock et al. (1995) and references therein.

The limit in the EUV band (entry 2) was inferred from the absence of the HeII 4686 line in the optical spectrum of A0620-00 (Marsh et al. 1994). The data covering 2400–4400Å (entries 4-14) were obtained with the HST Faint Object Spectrograph. The observations and data analysis are described by McClintock et al. (1995); their Figure 4b was averaged in 200Å intervals to give the fluxes listed in Table 1. At the longer wavelengths ($\gtrsim$ 3500Å), the contribution to the flux due to the K-star has been removed (McClintock et al. 1995). The estimated uncertainties in the fluxes, which are due primarily to systematic effects, are about 20%. The single flux measurement at 1775Å (entry 3), which was also obtained with the HST Faint Object Spectrograph, covers a much broader band (1350–2200Å). This measurement is compromised by uncertainties in the background and by the faintness of the source. Nevertheless, it is very likely that the flux at 1775Å given in Table 1 is in the range $-28.68 < \log F_\nu < -27.83$ (McClintock et al. 1995).

The 5000–10000Å fluxes (entries 15-20) were determined by Oke (1977). Again, the K-star's contribution to the flux ($\sim$ 55% at V) has been removed. The near-IR fluxes at J and K (entries 21-22) are primarily due to the K-star and provide only weak upper limits on the disk emission. The VLA radio limit at 6-cm wavelength (entry 23) does not constrain the models in question.

Several of the flux measurements summarized in Table 1 were obtained many years apart. Is source variability in quiescence a problem? Photometric optical data, collected from 1981 through the present, show that the intensity of the accretion disk is relatively stable (McClintock et al. 1995). Similarly, a stringent limit obtained in 1979 on the X-ray luminosity, $L_X < 10^{32}$ erg s$^{-1}$ (Long, Helfand, & Grabelsky 1981), is consistent with the 1992 ROSAT luminosity given above. We conclude that the data in Table 1 should not be seriously affected by variability.

### 3.2. Results of Modeling A0620-00

Figure 1 shows one of our models of A0620-00, corresponding to an inclination of $i = 70^o$ and black hole mass of $M = 4.4 M_\odot$. The viscosity parameter is $\alpha = 0.1$. In this model we have adjusted $\dot{M}$ in the inner two-temperature flow so as to fit the X-ray flux. This gave us $\dot{M} = 6.5 \times 10^{-5} \dot{M}_E = 6.3 \times 10^{-12} M_\odot \mathrm{yr}^{-1}$. We then used the same value of $\dot{M}$ for the outer thin disk and adjusted the inner radius of this disk $r_{in}$, or equivalently the inner orbital velocity $v_{in} = c \sin i / \sqrt{2 r_{in}}$ (cf. eq 3.2), so as to fit the optical data; the fit gave $v_{in} = 5000$ km s$^{-1}$. As can be seen, the model agrees with all the measured fluxes, and is consistent with the upper limits.

The dotted line in Fig. 1 shows the contribution to the spectrum due to the outer thin disk alone, and the dashed line corresponds to the emission from the inner hot flow. As expected, the optical and UV emission are dominated by the outer disk, but all the emission in X-rays is from the inner flow. The model predicts a substantial flux in hard X-rays out to about 1 MeV.

Figures 2a, 2b show models corresponding to the same values of $i$ and $M$, but with two other values of $\alpha$: 0.3 and 0.03. Since the X-ray flux is proportional to $\dot{M}/\alpha$ (§2.1), the values of $\dot{M}$ in these models have been scaled with respect to the model in Fig. 1 by $\alpha$. The parameter $v_{in}$ has been optimized in each model, but since the optical flux varies as $\sim \dot{M}/r_{in} \sim \dot{M} v_{in}^2$ (§2.2), we see that we have the approximate scaling $v_{in} \sim \alpha^{-1/2}$. The $\alpha = 0.3$ model is almost as good as the $\alpha = 0.1$ model, but the $\alpha = 0.03$ model can be definitely rejected. We thus conclude that the parameters of the model are uniquely determined to within a factor of $\sim 3$. The best fit $\alpha$ lies in the range $0.1 - 0.3$.

There are several noteworthy features of these models which we wish to emphasize.
1. The models are able to fit the optical and X-ray data simultaneously. Note that the UV flux is quite a bit lower than the optical flux, whereas the change from UV to X-rays is more modest. Thus the spectrum of A0620-00 must turn up rather suddenly between UV and X-rays. Our model produces this upturn quite nicely by having different components in the flow produce the optical and X-ray emission.
2. While we have adjusted $\dot{M}/\alpha$ so as to fit the X-ray flux, we have no control over the X-ray spectrum. It is satisfying that our predicted spectrum is consistent with the power law fits of the ROSAT data to within the $1\sigma$ error limits. We view this as an independent confirmation of the validity of the model. Similarly, the shape of the spectrum in the optical/UV band is consistent with the data (Figs. 1, 2a).

3. Although we treat $\alpha$ as a free fitting parameter, we find that the best value we obtain is $\alpha \sim 0.1 - 0.3$, which is similar to estimates of $\alpha$ in hot disks in CVs during outburst (e.g. Cannizzo 1993, Mineshige & Kusunose 1993). Note, however, that the $\alpha$ in our model refers to the quasi-spherical inner flow, which is quite a bit hotter than any CV disk.
4. The value of $\dot{M}$ we obtain is in the range $\sim 7 \times 10^{-12} - 2 \times 10^{-11} M_\odot \text{yr}^{-1}$. This is an interesting rate. As mentioned in §3.1, the total energy radiated by A0620-00 during outburst suggests that the source must store mass at a rate $\sim 3 \times 10^{-11} M_\odot \text{yr}^{-1}$ during quiescence (McClintock et al. 1983). It is gratifying that the accretion rate we estimate in our model is within a factor of a few of the estimated mass storage rate. The total mass transfer rate from the secondary is of course the sum of our estimated accretion rate and the mass storage rate.
5. By fitting $v_{in}$, the orbital velocity at the inner edge of the outer disk, as a free parameter we obtain values in the range $\sim 3000 - 5000$ km s$^{-1}$. As mentioned in §3.1, velocities up to 2800 km s$^{-1}$ have been inferred from the H$\alpha$ line profiles (Marsh 1994) and even these are lower limits. This is another independent confirmation of the model.
6. The model predicts a substantial luminosity in hard X-rays. None of the current X-ray telescopes appear to have enough sensitivity to detect this flux, but this would be an interesting test for future more sensitive telescopes.

In Fig. 3 we show models of A0620-00 for the alternate inclination of $i = 40°$ (Shahbaz et al. 1994), which corresponds to a black hole mass of $M = 12 M_\odot$. We are able to obtain reasonable models in this case as well, but only with lower values of $\alpha \sim 0.01$. Also, the mass accretion rate we derive is much lower, $\dot{M} \sim 10^{-12} M_\odot \text{yr}^{-1}$, which means that these models are less satisfactory according to point 4 above.

Finally, we show in Fig. 4 standard steady state thin disk models of A0620-00 where we assume that the thin disk extends all the way down to the marginally stable orbit at $r_{in} = 3$. The solid line shows a model where we adjusted $\dot{M}$ to pass through the optical data. This model disagrees quite severely with the UV measurement and the EUV upper limit. On the other hand, if we reduce $\dot{M}$ so as to fit the UV point (dashed line), the fit in the optical and X-rays is poor.

## 4. Modeling Other Soft X-Ray Transients

### 4.1. V404 Cyg

V404 Cyg is the most secure black hole candidate by virtue of its large mass function: f(M) = 6.08 ± 0.06 $M_\odot$ (Casares & Charles 1994). Its long orbital period, P = 155.3 hr, is extraordinary for an X-ray nova. (All other X-ray novae have periods < 19 hr.) The secondary is a K subgiant (or a stripped giant), and we adopt a distance of D = 3.5 kpc (Wagner et al. 1992; Shahbaz et al. 1994). Likely values for the remaining system parameters are summarized by Shahbaz et al. (1994): $i = 56°$, $M = 12\ M_\odot$, and $M_s = 0.7\ M_\odot$. An estimate of the projected velocity in the inner disk was obtained by measuring the H$\alpha$ profiles shown in Figure 9 of Casares et al. (1993): $v_{in} \geq 1140$ km s$^{-1}$.

V404 Cyg is significantly reddened, $A_V \sim 4.0$ mag. The dereddened optical and IR magnitudes are summarized in Table 3 of Casares et al. (1993). Also given in the same reference is the fraction of the total flux contributed by the accretion disk in the B, V and R bands. From these data we find: log $F_\nu$(B) = -26.09, log $F_\nu$(V) = -26.03, and log $F_\nu$(R) = -26.00 (with $F_\nu$ in units of erg s$^{-1}$ cm$^{-2}$ Hz$^{-1}$).

Pointed X-ray observations were made 3.5 yr after the 1989 May outburst using the ROSAT PSPC detector. For details see Wagner et al. (1994). We retrieved these data from the HEASARC data archive and analyzed them exactly in the manner described in §3.1 with one exception: V404 Cyg was relatively bright ($\approx$ 370 net source photons were detected in $1.6 \times 10^4$ s) and we were able to use $\chi^2$ model fitting instead of the more complicated likelihood fitting (cf. §3.1). We used an average of the data over the entire $1.6 \times 10^4$ s observation. (Note that the source was quite variable and declined in intensity by an order of magnitude during the observation; Wagner et al. 1994.) As with A0620-00, we fitted the data for V404 Cyg with a power law and fixed the interstellar column density at $N_H = 5.0 \times 10^{21}$ cm$^{-2}$ (Han & Hjellming 1992; Casares et al. 1993). The fitted pulse height data included channels 16-34 (nominally 0.7-2.5 keV). We varied only the photon spectral index and determined the following value: $\alpha_N = 1.25 \pm 0.30$. The corresponding X-ray flux is log $F_\nu$ = -29.75 ± 0.01, and the luminosity is $L_X \approx 1.1 \times 10^{33}$ erg s$^{-1}$ (0.7-2.4 keV).

The above values of the X-ray spectral parameters, flux and luminosity are the ones we have adopted for use in this paper. However, we also fitted a power law model to the data allowing the column density to be a free parameter, and we obtained the following results: $N_H = 2.1 \times 10^{22}$ cm$^{-2}$ and $\alpha_N = 4.0$ (-1.5, +1.9). (The uncertainties given here correspond to a confidence contour with $\Delta\chi^2 = 1$.) Our value for the column density is very comparable to the one obtained by Wagner et al. (1994); however, their power law slope is steeper, $\alpha_N = 7$. They do not give uncertainties.

Because of the absence of UV data for V404 Cyg, we are unable to determine $\dot{M}, \alpha$ and $v_{in}$ simultaneously from the data. Figure 5a shows two models of V404 Cyg, where based on our experience with A0620-00, we have tried $\alpha = 0.1$ and 0.3. In both cases, we are able to fit the optical and X-ray data reasonably well. The corresponding best-fit parameters are listed on the figures. The $\alpha = 0.3$ model appears to fit the "shape" of the optical data points somewhat better.

Because we do not have UV data to constrain the size of the outer thin disk, in principle the data on V404 Cyg can also be fit with a standard thin disk extending all the way down to the marginally stable orbit at $r_{in} = 3$. Fig. 5b shows such a model, where we have adjusted $\dot{M}$ so as to pass through the optical data points. The model gives an acceptable fit even to the X-ray flux. However, it makes a completely different prediction for the UV and EUV flux of the source compared to the models in Fig. 5a. Measurements in UV are very difficult, given the enormous extinction towards V404 Cyg. However, a limit on the EUV flux using the HeII 4686 line appears feasible (cf. the analogous limit obtained by Marsh et al. 1994 for A0620-00) and may provide a means of discriminating between the models.

Perhaps most promising as a test to distinguish between the two classes of models is an X-ray observation with ASCA, which has substantial effective area up to almost 10 keV. We simulated a 40 ks observation of V404 Cyg including a typical observed background spectrum. We used a power law spectrum with a photon index of $\alpha_N = 1.75$, which corresponds to our model shown in Fig. 5a. We adopted a small extraction aperture, 0.6 arcmin square, in order to suppress the background. For the sum of 2 SIS detectors we find the following numbers of net source counts as a function of energy: $1 - 2$ keV, $185 \pm 17$ cts; $2 - 3$ keV, $54 \pm 4$ cts; $3 - 4$ keV, $31 \pm 1$ cts; and $4 - 5$ keV, $16 \pm 2$ cts. Thus, the simulation predicts that ASCA would detect V404 Cyg to at least 5 keV if our model is correct. On the other hand, the simulation predicts that the pure disk spectrum (Fig. 5b) would definitely not be detected above 3 keV. We note that some evidence for a flat spectrum extending to 10 keV or more in quiescence has been reported (Mineshege et al. 1992). This evidence is not conclusive because of the large field of view (1.1 deg x 2.0 deg FWHM) used in making the observations. Nevertheless, this result underscores the need to observe V404 Cyg with the imaging capability of ASCA.

### 4.2. X-Ray Nova Mus 1991

Nova Mus has an orbital period of P = 10.4 hr and a mass function of f(M) = 3.01 $\pm$ 0.15 $M_\odot$, making it similar to A0620-00 (Orosz et al. 1995). An analysis of the ellipsoidal light curves suggests that the system is highly inclined to the line of sight. We therefore adopt a model with $i = 70°$ $M = 4.5$ $M_\odot$ and $M_s = 0.5$ $M_\odot$ (Orosz et al. 1995). A study of the H$\alpha$ emission line profiles suggests $v_{out} = 450 \pm 10$ km s$^{-1}$ and $v_{in} \geq 2000$ km s$^{-1}$ (Orosz et al. 1994).

As in the case of A0620-00 and V404 Cyg, the optical spectrum of Nova Mus has been decomposed into a component due to a K3-K5 dwarf and a component of comparable intensity due to an accretion disk. The spectrum of the accretion disk, corrected for the modest reddening of the source, E(B-V) = 0.29, is shown in Figure 4 of Orosz et al. (1995). Using this figure, we obtained the following accretion disk fluxes at four wavelengths in the continuum (i.e., excluding the strong emission lines): log $F_\nu$(5000Å) = -27.88, log $F_\nu$(5500Å) = -27.76, log $F_\nu$(6000Å) = -27.72, and log $F_\nu$(6500Å) = -27.72.

Pointed X-ray observations were made 1.2 yr after the nova outburst in 1991 January using the ROSAT PSPC detector. For details see Greiner et al. (1994). We obtained these data from the HEASARC data archive. A visual inspection of an image of the field revealed no evidence for an X-ray source at the location of the optical counterpart (van Paradijs 1995). However, there is a faint source located $\approx$ 2' east and somewhat north of the location of Nova Mus. In computing the background we avoided this source by selecting three circular background regions, which were centered respectively 3.5' due north, west and south of Nova Mus. These background apertures had radii of 2.5'. The source counts were extracted from a circular aperture of radius 1.0', which was centered on the source. We fixed the hydrogen column density at $N_H = 1.4 \times 10^{21}$ cm$^{-2}$ (Cheng et al. 1992; Heiles et al. 1981), and the photon spectral index at $\alpha_N = 3$ (cf., A0620-00 above) and derived the following upper limit on the (0.4 - 2.4 keV) X-ray flux: $\log F_\nu < -30.97$. The corresponding luminosity limit is $L_X < 6 \times 10^{31}$ erg s$^{-1}$. Both limits are at a $3\sigma$ level of confidence. For a flatter spectrum with $\alpha_N = 1$, the flux and luminosity limits are lower by about 10%.

In trying to model Nova Mus, we are hampered by the fact that we do not have an X-ray flux, only an upper limit. However, in contrast to V404 Cyg, the optical data do provide both a flux and a nominal temperature for the thin disk. (The temperature estimate is possible because the optical $\nu F_\nu$ has a peak inside the measured band.) We have therefore used the optical data and estimated both $\dot{M}$ and $r_{in}$ simultaneously. Our estimate of $\dot{M}/r_{in}$ is probably fairly accurate since it is obtained directly from the optical luminosity. However, since the optical data cover only a small range of frequencies, we cannot measure the temperature very well and so the estimates of the individual values of $\dot{M}$ and $R_{in}$ are probably not very accurate. Nevertheless, we have taken our best-fit parameters and calculated the emission we expect from the hot inner flow for the two standard values of $\alpha$: 0.1 and 0.3. Figure 6a shows the results. We see that the predicted X-ray flux is comfortably below the measured upper limits. Both models are therefore consistent with the available data.

As with V404 Cyg, we have also computed a standard thin disk model where the disk extends all the way down to $r_{in} = 3$. The spectrum is shown in Fig. 6b. This spectrum does not fit the shape of the optical data well. In addition, we see that it differs considerably from the two-zone models of Fig. 6a in the UV. Observations in UV should therefore be able to distinguish between the models.

## 5. Summary and Discussion

The quiescent state of soft X-ray transients (SXTs) has not been investigated in detail until now, largely because there has been very little observational data available. The observational situation has now improved for several SXTs, notably A0620-00. Detailed observations of A0620-00, both from the ground (Marsh et al. 1994, Orosz et al. 1994) and with the Hubble Space Telescope (McClintock et al. 1995), have provided a spectrum of the source covering almost a decade of frequency in the infrared, optical and ultraviolet bands. In addition, the source has been detected in X-rays with ROSAT (McClintock et al. 1995) and this has provided a flux measurement and an approximate spectral index in the energy range 0.4–2.4 keV.

The combined optical and X-ray data on A0620-00 are shown in Fig. 1. Even a cursory examination of this figure makes it clear that a standard thin accretion disk model cannot explain the observations. We see that the spectrum falls steeply in UV, and then apparently levels off in soft X-rays. Such a spectral shape is not what one expects from a thin disk. Therefore, at best, one can hope to explain either the X-ray data alone or the optical/UV data alone with a thin disk model.

The X-ray data on A0620-00 are incompatible with a thin disk because of the low luminosity ($L_X \sim 6 \times 10^{30}$ erg s$^{-1}$) and relatively high temperature ($kT \sim 0.2$ keV) observed. The low luminosity requires a very low accretion rate of $\dot{M} \sim 10^{-15} M_\odot$yr$^{-1}$ (or $\dot{m} \sim 10^{-6}$), if the disk extends down to the marginally stable orbit, but at such a low accretion rate the effective temperature is expected to be less than that observed (e.g. eq. 2.25 gives $T_{eff} \sim 6 \times 10^5$ K). Conversely, if we adjust $\dot{M}$ to fit the temperature, the implied X-ray luminosity will be much too large. These difficulties are illustrated by Fig. 4.

The optical/UV data by themselves are compatible with a thin disk model. Indeed, McClintock et al. (1995) commented that the optical spectrum of A0620-00 resembles that of low-$\dot{M}$ cataclysmic variables (CVs) rather closely. However, this resemblance is not expected and presents a serious problem. In CVs, the disk only extends down to the surface of the accreting white dwarf, which is at a radius of $\sim 5000$ km, or $r \sim$ few $\times 10^3$ in Schwarzschild units (see eq. 2.15 for the distinction between the physical radius $R$ and the dimensionless radius $r$). However, most SXTs are believed to have accreting black holes (some are accreting neutron stars, cf. van Paradijs & McClintock 1995), and the accretion flow must extend all the way down to the horizon. The optical/UV data on A0620-00 cannot be fit with any thin disk model where the inner radius $r_{in}$ is $\sim 1$. Instead, as we show in this paper, the data are fitted well with a thin disk with $r_{in} \sim 3000 - 5000$. This means that either these systems have no accretion at all inside of $r_{in} \sim$ few $\times 10^3$, in which case it is hard to understand where the X-ray luminosity is coming from, or the accretion flow inside of this radius must occur in a form very different from the standard thin disk.

Motivated by this, we propose for A0620-00 and two other SXTs, V404 Cyg and Nova Mus, a model consisting of two distinct zones of accretion. On the outside, extending from $r_{in} \sim 3000 - 5000$ out to the outermost radius $r_{out}$ of the accretion flow, we have a standard thin accretion disk. The emission from this component dominates in the infrared, optical and UV. For $r < r_{in}$, however, there is no thin disk. Instead, we have a completely different kind of flow, an advection-dominated flow of the form discussed in several recent papers (Narayan & Yi 1994, 1995ab, Abramowicz et al. 1995, Chen et al. 1995). In this flow, the accreting gas is very hot, almost at virial temperature, and optically thin. The emission from this region has a number of different components. First, there is a primary component due to synchrotron emission by the thermal electrons in the ambient magnetic field. (We fix the strength of the field by assuming that the magnetic pressure is equal to a tenth of the equipartition value, but the exact choice is not important). The synchrotron emission is primarily in the optical but since its amplitude is a small fraction of the radiation coming from the outer thin disk this radiation is not distinguishable as a separate component in the combined spectrum. Some of the synchrotron photons are Compton upscattered. Those photons which are singly scattered provide a significant luminosity in $0.1 - 1$ keV X-rays. This part of the spectrum explains the X-ray flux observed in A0620-00 (McClintock et al. 1995, §3.1) and V404 Cyg (Wagner et al. 1994, §4.1). Finally, there is emission in hard X-rays as a result of bremsstrahlung emission plus multiply Compton-scattered synchrotron photons. There are no detections as yet in this region of the spectrum.

As Figs. 1–6 show, the models we present are quite successful in explaining all the available data on the three sources. From our model fits we determine three parameters: the viscosity parameter $\alpha$ in the hot flow, the mass accretion rate $\dot{M}$, which we assume to be the same in the outer and inner zones, and the radius of the inner edge $r_{in}$ of the outer thin disk. Our fits show that $\alpha$ probably lies in the range $\sim 0.1 - 0.3$. Note that this refers to the hot inner flow and we do not have any constraint on the viscosity parameter $\alpha_t$ in the outer thin disk. We deduce a mass accretion rate $\sim 10^{-11} M_\odot \mathrm{yr}^{-1}$ in A0620-00 and Nova Mus, and a higher rate $\sim$ few $\times 10^{-10} M_\odot \mathrm{yr}^{-1}$ in V404 Cyg. Since V404 Cyg has a sub-giant secondary whereas the others have main sequence companions, a larger $\dot{M}$ in this source is plausible; moreover, our estimate for $\dot{M}$ is in agreement with the predictions made by King (1993). The third parameter, $r_{in}$, turns out to be surprisingly similar in all three sources, lying in the range 3000-5000 Schwarzschild radii.

Although the three parameters, $\alpha$, $\dot{M}$, $r_{in}$, are used purely as fitting parameters in our models, the values that we derive for them are reasonable. Firstly, the value of $\alpha \sim 0.1 - 0.3$ which we obtain is similar to the value deduced for CVs in outburst (e.g. Smak 1983, 1994, Cannizzo 1993, Mineshige & Kusunose 1993). To the extent that our hot flows are closer in spirit to the outburst state rather than the quiescent state of CVs, it is satisfactory that our $\alpha$ is closer to the outburst value (which is $\alpha_t \gtrsim 0.1$) than the quiescent value (which is $\alpha_t \sim 0.01$). Secondly, the $\dot{M} \sim 10^{-11} M_\odot \mathrm{yr}^{-1}$ which we estimate for A0620-00 is quite similar to the mass storage rate of $\sim 3 \times 10^{-11} M_\odot \mathrm{yr}^{-1}$ deduced for this system from the total energy output in outburst (McClintock et al. 1983, cf. §3.1). Of course, there is no rigorous argument which says that these two rates should be similar. However, we feel that it is much more natural for the rates to be comparable than for them to differ by orders of magnitude. In this sense, the accretion rate we derive is satisfactory. Finally, the value of $r_{in}$ we deduce in A0620-00 corresponds to an orbital velocity of $3000 - 5000$ km s$^{-1}$ at the inner edge of the disk. From the H$\alpha$ line profiles of the source, it appears that the velocity at the inner edge is at least 2800 km s$^{-1}$ (Marsh et al. 1994). Thus, the value of this parameter again seems to agree with independent observational information.

In the case of A0620-00, where we have the most complete data, our models fit the observations remarkably well. In V404 Cyg and Nova Mus there are no infrared or UV data, and in Nova Mus there is not even an X-ray detection. Because of this, our models of these sources are less constrained, and it appears that even a pure thin disk model may be marginally consistent with the data. Our preferred two-zone models for these sources, shown in Figs. 5a, 6a, have a significantly different spectrum than the thin disk models, shown in Figs. 5b, 6b. The differences are especially striking in UV, EUV and X-rays. It would therefore be extremely interesting to extend observations of Nova Mus into the UV with the Hubble Space Telescope, to set limits on the EUV both for V404 Cyg and Nova Mus via observations of He II $\lambda 4686$, and to look for hard ($\sim 2 - 5$ keV) X-rays from V404 Cyg with ASCA.

The two-zone structure of the accretion flow is treated as just an ansatz in this paper, and we have used the transition radius $r_{in}$ as a free fitting parameter. Interestingly, the values of $r_{in}$ that we obtain for the three sources from our best-fit models all correspond to the radius at which the disk achieves a surface temperature of $10000 - 15000$ K. It is not clear if this is a coincidence or a hint as to the nature of the instability which causes the disk to evaporate. The idea of a thin disk evaporating to form a corona has been discussed recently by Meyer & Meyer-Hofmeister (1994) in the context of CVs. It would be very interesting to apply their mechanism to SXTs and to see if the model can reproduce the values of $r_{in}$ which we have obtained by fitting the spectra.

The fact that accretion flows around black holes (and neutron stars) must involve hot flows rather than the standard thin disk has been known for a long time. Accreting black holes frequently radiate in hard X-rays and gamma-rays, with the luminosity in these bands often being considerable (e.g. Sunyaev et al. 1993). The high temperature implied by this radiation is just not compatible with a thin disk, which is usually relatively cool (see eq 2.20). Until recently, the only hot solution known was the two-temperature solution discovered by Shapiro et al. (1976) and developed by later authors (e.g. Kusunose & Takahara 1989, Wandel & Liang 1992, Luo & Liang 1994, Melia & Misra 1993). Unfortunately, this solution is known to be thermally unstable (Piran 1978), and therefore its relevance to accretion flows is unclear.

The advection-dominated solution which we have used for our hot inner flow is similar to the Shapiro et al. (1976) solution in that it is hot, optically thin, and two-temperature. However, it is both thermally and viscously stable (Abramowicz et al. 1995, Narayan & Yi 1995b). Why does our flow not suffer from the thermal instability? The reason is that the flow is "advection-dominated." This means that most of the energy which is released by viscosity is retained in the gas, and only a small fraction is radiated. In a sense, the gas *is* thermally unstable. However, even as it heats up under the effect of the instability it continues to accrete inwards until it asymptotically achieves a nearly self-similar virial configuration. The self-similar solution of advection-dominated flows described in Narayan & Yi (1994) and employed in this paper (§2.1) describes the situation in a convenient analytical form. It should be mentioned that prior to the recent work on optically thin solutions, a branch of advection-dominated solutions had been discovered by Abramowicz et al. (1988) for optically thick flows at very high $\dot{M}$. These solutions have been further investigated by Honma, Matsumoto & Kato (1991) and Chen & Taam (1993). The optically thin solutions discussed here represent a different branch of solutions. These solutions occur at much lower $\dot{M}$ and are much hotter.

The fact that our hot inner flow is advection-dominated explains the low X-ray luminosities of quiescent SXTs. Since the same mass accretion rate $\dot{M}$ flows through the outer thin disk and the inner hot flow in our models, naively one would have expected the luminosity from the inner flow to exceed that of the outer disk by a factor $\sim 0.1 r_{in} \sim$ few $\times 10^2$, assuming an efficiency of 10% for the inner flow. Instead, the X-ray luminosities of all three SXTs we have modeled are *less* than the corresponding optical/UV luminosities from their outer thin disks. This is because the inner flow has a very low efficiency, $\sim 10^{-4} - 10^{-3}$, so that more than 99.9% of the viscous energy that is dissipated in the inner region is transported into the black hole.

We believe that an exactly analogous situation is present in the source Sgr A* at the center of our Galaxy (Narayan, Yi & Mahadevan 1995). We have developed a model for this source where a $7 \times 10^5 M_\odot$ black hole accretes at a fairly large rate of $\sim 10^{-5} M_\odot \text{yr}^{-1}$. Yet, the total luminosity is only $\sim 10^{37}$ erg s$^{-1}$. The explanation again is that most of the accretion energy is lost into the supermassive black hole and only $\sim 10^{-4}$ of the energy is radiated.

These studies provide a warning that it is important to allow for the low efficiency of advection-dominated flows when estimating the mass accretion rates of accreting black holes. When an accreting black hole is seen to emit a substantial fraction of its luminosity in hard X-rays or gamma-rays, it is plausible to assume that the inner regions of the flow involve the new advection-dominated solution we have used in this paper, since this is currently the only stable accretion solution we are aware of that achieves the high temperatures needed to produce hard radiation. Since advection-dominated flows are, by definition, *inefficient* emitters of radiation, the real $\dot{M}$ in such systems is likely to be higher than the nominal rate one usually assumes, $\dot{M}_\text{nominal} \sim L/0.1c^2$, where $L$ is the observed luminosity. In some cases, such as Sgr A* (Narayan et al. 1995) or quiescent SXTs (this paper), $\dot{M}$ could be higher than $\dot{M}_\text{nominal}$ by as much as a factor of $10^3 - 10^4$.

It is instructive to compare the quiescent X-ray emission of a black hole transient and a neutron-star transient, such as the well-studied system Cen X-4. The quiescent X-ray luminosity of Cen X-4 is $\simeq (1.5 - 4.2) \times 10^{33}$ ergs s$^{-1}$ (van Paradijs et al. 1987), which is $\sim 200 - 1000$ times the quiescent X-ray luminosity of A0620-00. If we assume that the mass accretion rates of the two systems are comparable, then their very different luminosities can be explained by the low efficiency we predict for accretion through the event horizon of a black hole ($\sim 10^{-3} - 10^{-4}$), and the 100 - 1000 times greater efficiency expected for accretion onto the surface of a neutron star (Narayan & Yi 1995b). The low X-ray luminosity of A0620-00 (and other black hole transients) is an important prediction of our model. On the other hand, the much larger luminosity of neutron star transients is less specific to our model, and is consistent with a wide range of accretion models.

All optically thin advection-dominated flows are extremely hot, with electron temperatures $T_e \sim 10^{9.5} - 10^{10}$ K. The temperatures at the very low accretion rates considered in this paper are in fact even a little greater than $10^{10}$ K, whereas at somewhat higher $\dot{m}$ the temperature tends to be closer to $10^{9.5}$ K (cf. Narayan & Yi 1995b). The spectra corresponding to these flows are invariably very hard, but not necessarily a single power law, as indicated by our models in Figs. 1–5. Detection of hard X-ray emission in quiescent SXTs will be a valuable confirmation of our model. Incidentally, the mass transfer instability model of SXT outbursts (Hameury et al. 1986, 1990) requires substantial irradiation of the secondary by hard radiation from the accretor. This is provided naturally in the quiescent models presented here.

Another feature of the advection-dominated solutions is that the X-ray luminosity is not proportional to $\dot{M}$, but tends to be more nearly proportional to $\dot{M}^2$. This is because the dominant cooling is usually via Comptonization of synchrotron photons, or occasionally through bremsstrahlung radiation, and these processes vary as the square, or even a higher power, of the density. This effect provides a plausible explanation for the difference in luminosity between V404 Cyg and the other two SXTs, A0620-00 and Nova Mus. The X-ray luminosity of V404 Cyg is almost $10^3$ times greater than that of the other two sources, which would normally imply a $10^3$ times larger $\dot{M}$ in this source. In contrast, in our model, the $\dot{M}$ of V404 Cyg is only $\sim 30$ times greater than in the other two sources, and yet because of the strong dependence of luminosity on accretion rate, we are able to reproduce the nearly factor of $10^3$ difference in luminosity. In an analogous way, we have explained (Narayan et al. 1995) the factor of $\sim 50$ X-ray variability of Sgr A* with only a factor of $\sim 4$ variation in $\dot{M}$ in that source.

Prior to this work, most of the theoretical effort on SXTs was devoted to modeling the *outbursts* of these binary systems. The favored model presently is the disk instability model (Mineshige & Wheeler 1989, Huang & Wheeler 1989, Canizzo 1994, Lasota 1995) where the disk undergoes a thermal limit cycle similar to that believed to produce dwarf nova eruptions in CVs (Papaloizou, Faulkner & Lin 1983, Mineshige & Osaki 1983, Meyer & Meyer-Hofmeister 1984, Smak 1984). If the model we have proposed here for quiescent SXTs is correct, then some of the details of the disk instability model may need to be revised since the models described so far in the literature automatically assume that the thin disk in a quiescent SXT extends all the way down to the marginally stable orbit at $r = 3$. Our disks are truncated at a very much larger radius, $r_{in} \sim 3000$, and this may lead to some quantitative changes in the predictions of the model. On the other hand, the disk instability model predicts that the accretion rate $\dot{M}$ varies with radius in the quiescent state of SXTs. For lack of detailed information on the profile of $\dot{M}(R)$ we have assumed a constant $\dot{M}$ in the models described in this paper. We do not believe this simplification is very serious. Nevertheless, future models should merge our ideas with the disk instability model and make use of a self-consistent $\dot{M}(R)$.

Another topic for future research is the nature of the flow during the outburst. The luminosity of an SXT during outburst approaches the Eddington limit, which implies that the mass accretion rate $\dot{M}$ onto the black hole increases by several orders of magnitude from the $\dot{M}$ values we have estimated for the quiescent state. The increased accretion is presumably fed by mass which has accumulated in the outer disk. How does this mass reach the black hole? It seems reasonable to assume that the extra mass initially evaporates into the corona and flows into the black hole as a hot advection-dominated flow with enhanced $\dot{M}$. Then, later on, as $\dot{M}$ builds up to a very large value, the thin disk itself probably extends all the way down to the black hole. If this picture is qualitatively correct, we expect the spectrum to be hard during the early hot stages of the outburst, but to soften later at the peak of the outburst when the entire flow is via a cool thin disk. Such a spectral evolution was indeed observed during the 1975 outburst of A0620-00 (Ricketts, Pounds, & Turner 1975, Carpenter et al. 1976) and the 1991 outburst of Nova Mus (Lund 1993). Similarly, during the decline from peak, a hole probably forms at the center of the thin disk and we imagine that the hole slowly grows as the disk shrinks back to its quiescent state. Therefore, as the luminosity declines from the peak, we would expect the soft radiation to drop quickly and the hard component to persist for a somewhat longer time. Nova Mus exhibited this behavior during its decline (Gilfanov et al. 1993).

One of the features of our hot advection-dominated inner flow is that it consists of a two-temperature plasma, where the ions are almost at the virial temperature ($T_i \sim 10^{12.5}/r$ K) while the electrons are significantly cooler ($T_e \sim 10^{10}$ K). A long-standing uncertainty in this subject is whether or not such a two-temperature plasma can be maintained for any length of time. Begelman & Chiueh (1988) analyzed a particular mechanism whereby the ions and electrons may come into thermal equilibrium, and it is believed that hot plasmas may well have several other energy transfer channels that could induce thermal equilibrium between the two species. Rees et al. (1982) have made the interesting observation that the issue of whether or not such equilibrium between ions and electrons actually happens will not be decided by theoretical arguments but may have to be settled by observations of astrophysical sources. The success we have had with our two-temperature models of quiescent SXTs and Sgr A$^*$ (Narayan et al. 1995) could perhaps be viewed as an observational confirmation that nature does allow hot two-temperature plasmas in low-$\dot{m}$ accretion flows around black holes.

Another very interesting point is that the SXT models described here, as well as the model of Sgr A$^*$ described in Narayan et al. (1995), require a central horizon through which all the advected energy disappears before it can be radiated. This means that our models require not only a relativistic potential to produce the relativistic temperatures, but also a true event horizon since this is the feature which allows our models to have a very low radiative efficiency in the hot inner flow. If instead of a horizon, we had a regular star with a surface, all the advected energy would ultimately be radiated from the stellar surface and we would see the energy somewhere in the spectrum. The success we have had with our models could thus be construed as a "proof" that the central stars in A0620-00, V404 Cyg, Nova Mus and Sgr A$^*$ are truly black holes with horizons.

We thank Jean-Pierre Lasota for stimulating discussions on the limitations of current models of quiescent SXTs, Lucas Macri for help in simulating an ASCA observation of V404 Cyg, and Julian Krolik for comments. This work was supported in part by NASA Grant NAG 52837, and has made use of data obtained through the High Energy Astrophysics Science Archive Research Center Online Service, provided by the NASA-Goddard Space Flight Center.

TABLE 1
A0620-00: Spectrum of the Accretion Disk

| Entry Number | Wavelength (Å) | Log $\nu$ (Hz) | log $F_\nu$ (erg cm$^{-2}$s$^{-1}$Hz$^{-1}$) | References |
|---|---|---|---|---|
| 1 | 12.4 | 17.383 | $-31.17$ | 1,2 |
| 2 | 124 | 16.476 | $< -28.51$ | 3 |
| 3 | 1775 | 15.228 | $-28.08$ | 2 |
| 4 | 2400 | 15.097 | $-27.31$ | 2 |
| 5 | 2600 | 15.062 | $-27.17$ | 2 |
| 6 | 2800 | 15.030 | $-27.15$ | 2 |
| 7 | 3000 | 15.000 | $-27.11$ | 2 |
| 8 | 3200 | 14.972 | $-26.98$ | 2 |
| 9 | 3400 | 14.945 | $-26.88$ | 2 |
| 10 | 3600 | 14.921 | $-26.85$ | 2 |
| 11 | 3800 | 14.897 | $-26.82$ | 2 |
| 12 | 4000 | 14.875 | $-26.79$ | 2 |
| 13 | 4200 | 14.854 | $-26.79$ | 2 |
| 14 | 4400 | 14.833 | $-26.78$ | 2 |
| 15 | 5000 | 14.778 | $-26.75$ | 4 |
| 16 | 6000 | 14.699 | $-26.71$ | 4 |
| 17 | 7000 | 14.632 | $-26.67$ | 4 |
| 18 | 8000 | 14.574 | $-26.65$ | 4 |
| 19 | 9000 | 14.523 | $-26.62$ | 4 |
| 20 | 10000 | 14.477 | $-26.60$ | 4 |
| 21 | 12500 | 14.380 | $< -25.96$ | 5 |
| 22 | 22000 | 14.134 | $< -25.99$ | 5 |
| 23 | $6.17 \times 10^8$ | 9.687 | $< -26.85$ | 2 |

References,–(1) This work; (2) McClintock et al. 1995; (3) Marsh et al. 1994; (4) Oke 1977; (5) Shahbaz et al. 1994.

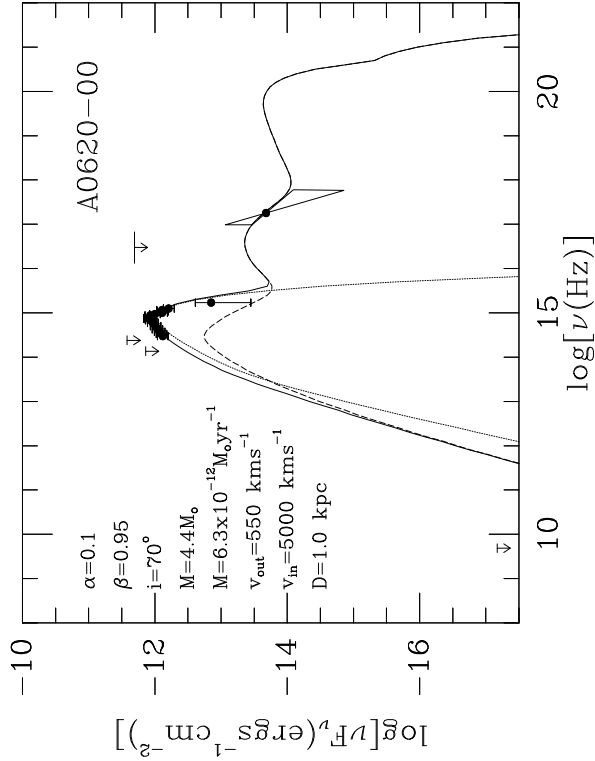

Fig. 1. The filled circles show measured fluxes of A0620-00 plotted as $\nu F_\nu$ vs $\nu$. The ROSAT point at $\nu \sim 10^{17}$ Hz ($h\nu \sim 1$ keV) is shown along with a "bow-tie" to indicate the range of spectral slopes allowed by the data. The downward pointing arrows represent the various upper limits in radio, infrared and EUV described in the text. The lines show the spectrum predicted by our model, where we have taken an inclination of $i = 70^o$, black hole mass $M = 4.4 M_\odot$, and viscosity parameter $\alpha = 0.1$. The parameters of the model are indicated on the left. The dotted line shows the contribution to the spectrum from the outer thin disk, the dashed line corresponds to the inner hot flow, and the solid line is the combined spectrum. Note that the combined spectrum fits all the data points well and is consistent with the upper limits.

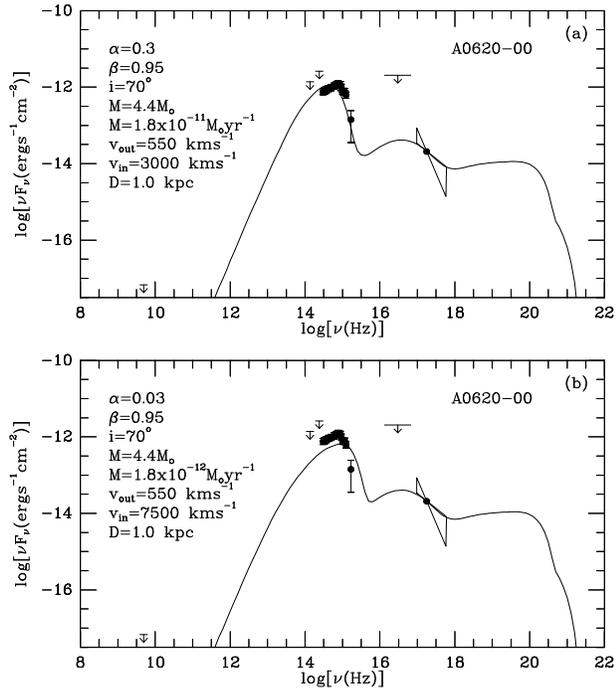

Fig. 2. (a) Similar to Fig. 1, but for $\alpha = 0.3$. Only the combined spectrum is shown. This model of A0620-00 fits the infrared data less well than the model in Fig. 1, but does better in UV. We consider the range of good models to lie between $\alpha = 0.1-0.3$. (b) A model of A0620-00 corresponding to $\alpha = 0.03$. We consider this model unacceptable.

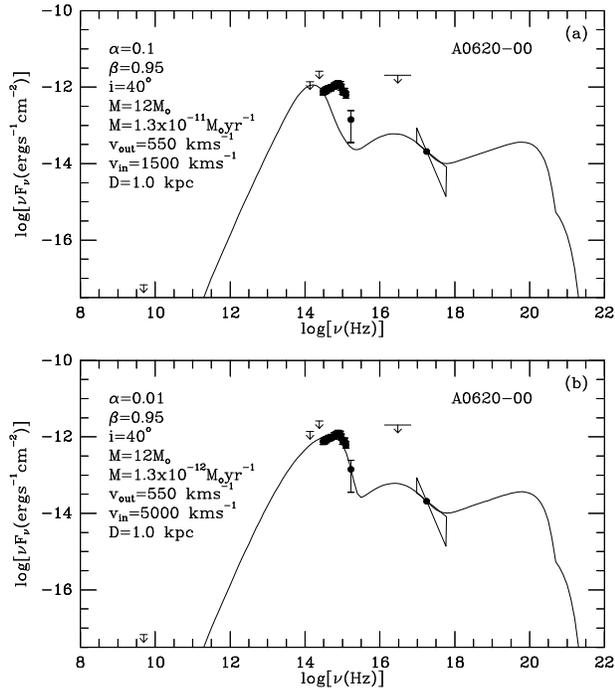

Fig. 3. Models of A0620-00 corresponding to an inclination $i = 40^o$. Here the best model corresponds to $\alpha = 0.01$.

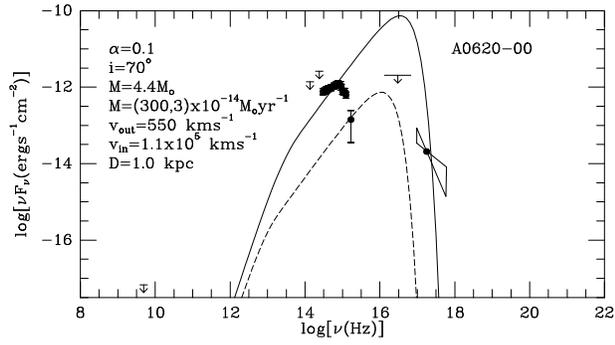

Fig. 4. Standard thin disk models of A0620-00, where the disk is allowed to extend down to the marginally stable orbit, $r_{in} = 3$. For no choice of $\dot{M}$ does the model fit all the data points.

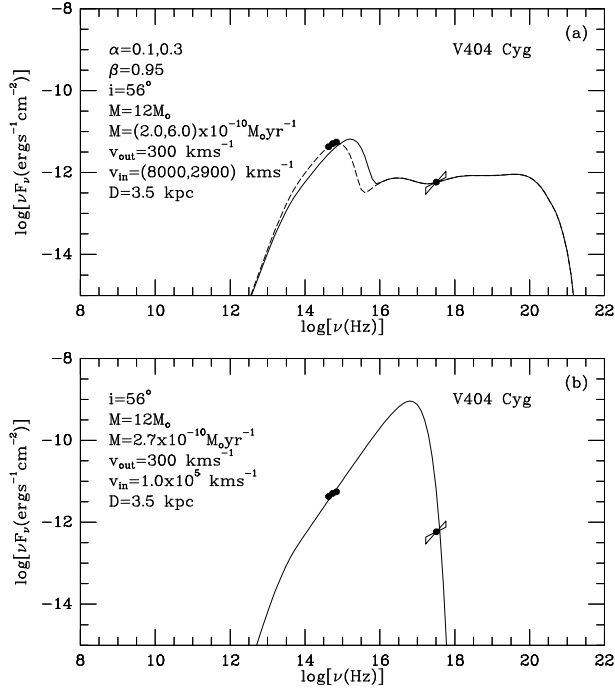

Fig. 5. (a) Two models of V404 Cyg corresponding to $\alpha = 0.1$ (solid line) and 0.3 (dashed line). The parameters of the models are shown on the left. The $\alpha = 0.3$ model appears to be superior. (b) The best-fit thin disk model, where the disk is allowed to extend down to the marginally stable orbit, $r_{in} = 3$. This model can be distinguished from the models in (a) by observations in 2–5 keV X-rays or by attempting to set a limit in the EUV band as in A0620-00.

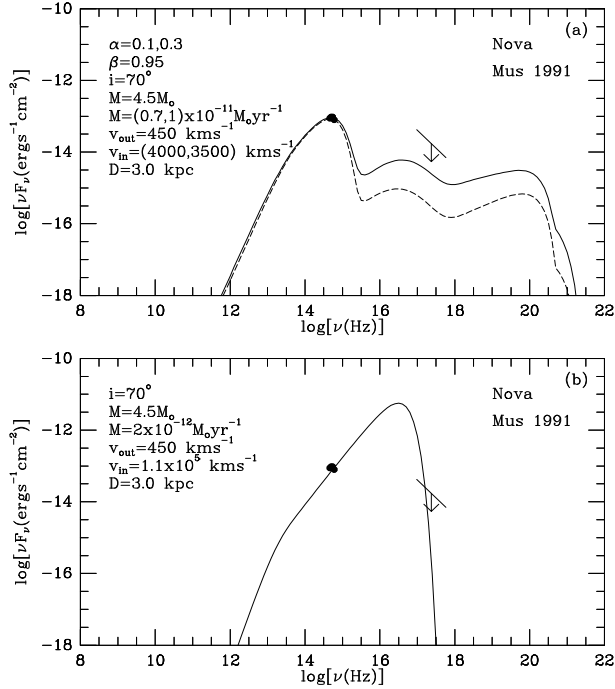

Fig. 6. (a) Two models of X-ray Nova Mus 1991 corresponding to $\alpha = 0.1$ (solid line) and 0.3 (dashed line). Both models are consistent with the optical data and the X-ray upper limit. (b) The best-fit thin disk model, where the disk is assumed to extend down to the marginally stable orbit, $r_{in} = 3$. This model can be easily distinguished from the models in (a) by observations in UV.